\documentclass[%
 aip,
 rsi,
amsmath,amssymb,
reprint,
floatfix
]{revtex4-1}

\usepackage{graphicx}
\usepackage{dcolumn}
\usepackage{bm}

\usepackage[utf8]{inputenc}
\usepackage{mathptmx}
\usepackage{multirow}
\usepackage{braket}

\usepackage[english]{babel}
\usepackage{xcolor}
\colorlet{RED}{red}
\colorlet{BLUE}{blue}
\usepackage{dcolumn}
\usepackage{bm}
\usepackage[version=4]{mhchem} 
\usepackage{acronym}
\usepackage{adjustbox}
\usepackage{float}
\usepackage{tikz}
\usepackage{microtype} 
\usepackage{algorithm}
\usepackage{algpseudocode}

\usetikzlibrary{calc,shapes.geometric,decorations.pathmorphing,patterns}

\definecolor{background-color}{gray}{0.98}
\usepackage[margin=2.3cm,bmargin=1cm,footnotesep=1cm]{geometry}

\begin{document}

\title{Coupled cluster
downfolding techniques: a review of existing applications in classical
and quantum computing for chemical systems
}

\author{Nicholas P. Bauman}
\email{nicholas.bauman@pnnl.gov}
\affiliation{%
  Pacific Northwest National Laboratory, Richland, Washington, 99354, USA
 }

\author{Bo Peng}
\email{peng398@pnnl.gov}
\affiliation{%
  Pacific Northwest National Laboratory, Richland, Washington, 99354, USA
}

\author{Karol Kowalski}
\email{karol.kowalski@pnnl.gov}
\affiliation{%
  Pacific Northwest National Laboratory, Richland, Washington, 99354, USA
 }

\begin{abstract}
  In this manuscript, we provide an overview of the recent developments of the coupled cluster (CC) downfolding methods, where the ground-state problem of a quantum system is represented through effective/downfolded Hamiltonians defined using active spaces. All CC downfolding techniques discussed here are derived from a single-reference  exponential ansatz for the ground-state problem. We discuss several extensions of the non-Hermitian and Hermitian downfolding  approaches to the time domain and the so-called quantum flows. We emphasize the important role of downfolding formalisms in transitioning chemical applications from noisy quantum devices to scalable and error-corrected quantum computers. 
\end{abstract}

\maketitle

\section{Introduction}
The coupled cluster (CC) theory
\cite{coester58_421,coester60_477,cizek66_4256,paldus72_50,purvis82_1910,arponen83_311,bishop1991overview,jorgensen90_3333,paldus07,crawford2000introduction,bartlett_rmp}
has evolved into one of the most accurate formulations to describe the correlation effects in chemistry,\cite{paldus07,crawford2000introduction,bartlett_rmp} 
material sciences, and physics.
\cite{arponen1987extended1,arponen83_311,arponen1987extended,arponen1991independent,arponen1993independent,robinson1989extended,arponen1991holomorphic,emrich1984electron,kummel2001post,PhysRevC.69.054320,PhysRevLett.92.132501,PhysRevLett.101.092502}
Although the CC formalism originates in the Linked Cluster Theorem,
\cite{brandow67_771,lindgren12}
it has been successfully extended to describe excited states, properties, and time evolution of the system.\cite{monkhorst77_421,jorgensen90_3333,arponen83_311,bartlett89_57,stanton93_7029,xpiecuch,krylov2008equation,
kknascimento2016linear,nascimento2017simulation,vila2022real,kvaal2012ab,pedersen2019symplectic,sato2018communication,pedersen2019symplectic,sato2018communication}
Over the last few decades, a significant effort has been exerted to address the steep scaling of canonical  CC formulations and apply them to realistic chemical processes. Parallel computing, especially with recently developed exascale computing architectures, has extended the applicability of conventional CC methods, but only modestly before encountering prohibitive costs once more. As a result, there has been much development in recent years on new reduced-scaling approaches for classical and quantum computing paradigms to push the envelope of the system sizes tractable by CC formalisms.

Mathematically rigorous formulations for reducing the dimensionality/cost of quantum formulations are urgently needed to shift the envelope of system-size tractable by accurate many-body formulations in chemistry, material sciences, and physics.  Among the most successful formulations, one should mention local coupled cluster (CC) formulations, various partitioning and incremental schemes, and embedding methods.
\cite{neese2009efficient,neese2009efficient2,riplinger2013efficient,Neese16_024109,pavosevic2016,pavosevic2017,rolik2013efficient,nagy2019approaching,eriksen2021incremental,li2010improved}
These approaches are driven by various design principles from the locality of correlation effects in the wave function approaches to properties of self-energy in correlated systems. Thanks to these formulations, significant progress has been achieved in describing correlation effects in large molecular systems allowing for simulations based on the utilization of modest computational resources.

The dimensionality reduction techniques  also play a crucial role in enabling the early stages of quantum computing driven by noisy intermediate-scale quantum devices (NISQ). This is associated with the reduction  of the qubits required to represent the quantum problem of interest. As an illustration, one should mention several techniques developed to take full advantage of the ubiquitous Variational Quantum Eigensolvers (VQE) approach 
\cite{peruzzo2014variational,mcclean2016theory,romero2018strategies,PhysRevA.95.020501,Kandala2017,kandala2018extending,PhysRevX.8.011021,huggins2020non,ryabinkin2018qubit,cao2019quantum,ryabinkin2020iterative,izmaylov2019unitary,lang2020unitary,grimsley2019adaptive,grimsley2019trotterized,cerezo2021variational,mcardle2020quantum,bharti2022noisy}
in addressing problems beyond the situation where few electrons are correlated. 

In the context of the development of quantum algorithms for quantum chemistry, the main goal of dimensionality reduction methods is to provide a mathematically rigorous way of representing interdependencies between static and  dynamical correlation effects. However, while the inclusion of static effects can be achieved for small-size systems on currently available quantum hardware, 
much needed dynamical correlation effects, usually manifesting in a  large number of fermionic degrees of freedom (amplitudes) characterized by small values, are beyond the reach of current quantum technologies. 

The recently introduced downfolding techniques based on the double unitary coupled cluster Ansatz (DUCC) \cite{bauman2019downfolding,downfolding2020t,metcalf2020resource,bauman2019quantumex,bauman2020variational,baumanpengGFDUCC,bauman2022coupled,bauman2022coupled2c} provide one of the solutions to the above-mentioned problem. The DUCC formalism offers a special representation of the ground-state  wave function  that, in analogy to single-reference sub-system embedding sub-algebras (SES-CC),
\cite{safkk,sssc}
allows one to construct effective Hamiltonians that integrate out all out-of-active-space degrees of freedom usually identified with dynamical amplitudes. 
Although the effective Hamiltonian formulations have a long history in quantum history and physics, especially in dealing with strong correlation effects, there are notable distinct features of the DUCC and SES-CC formalisms:
(1) both formulations are embedded in the single-reference language employing a straightforward definition of the excitation domain (i.e., wave function parameters) in the vain of single-reference formulations, and (2) the possibility of describing  a quantum problem in the form of quantum flows, i.e., coupled small-dimensionality eigenvalue problems. In this way, one can probe large sub-spaces of the Hilbert space without unrealistic quantum resource demands.
Since the eigenvalue problems involved in the quantum flow represent physically well-defined problems (defined by the corresponding effective Hamiltonians and density matrices), the quantum flow formulation naturally lent itself to capture possible sparsity characterizing the quantum system.  

This paper provides a compact overview of the main development threads originating in the single-reference SES-CC formulation (Section 2.1) and its unitary extension (Section 2.2). Section 3 introduces and discusses the basic tenets of quantum flows. The extension of the CC downfolding methods to the time domain and Green's function formalism is discussed in Sections 4 and 5.
Finally, Section 6 discusses applications of the downfolding formalisms.

\section{Theory}
The SES-CC and DUCC formulations have been amply discussed in recent papers (see Refs.~\cite{safkk,bauman2019downfolding,bauman2022coupled}). Here we overview only the salient features of these approaches. While the SES-CC technique forms the basis for non-Hermitian downfolding, the DUCC expansions provide its Hermitian formulations. In both cases, the ensuing downfolding procedures are encoded in the properties of exponential ansatzes for the ground-state wave functions $|\Psi\rangle$: 
\begin{equation}
|\Psi\rangle = e^T |\Phi\rangle \;,
\label{cc}
\end{equation}
for non-Hermitian formalism given by standard SR-CC expansion, and 
\begin{equation}
|\Psi\rangle = e^{\sigma_{\rm ext}}
e^{\sigma_{\rm int}}|\Phi\rangle \;,
\label{ducc}
\end{equation}
for the Hermitian downfolding defined by the DUCC Ansatz. In these equations, $|\Phi\rangle$ is the so-called reference function usually identified with the Hartree--Fock determinant, $T$ is the SR-CC cluster operator, and $\sigma_{\rm ext}$ and 
$\sigma_{\rm int}$ are the anti-Hermitian external and internal cluster operators ({\it vide infra}).

Both types of downfolding lead to many-body forms of effective or downfolded Hamiltonians acting in the appropriate active spaces. Although effective Hamiltonian formulations have a long history in electronic structure theory, especially in treating strong correlation effects, the present methods have several unique features compared to the multi-reference effective Hamiltonian approaches. Among the most distinct, one should mention: (1) the possibility of developing effective Hamiltonian formalisms using a very simple single-reference language to define the manifold of excitations used to construct downfolded Hamiltonians and (2) the concept of the quantum flows (QF), which boils down to coupling downfolding procedures corresponding to various active spaces. The former formalism allows for sampling large sub-spaces of the Hilbert space using reduced-dimensionality  eigenvalue problems. The QF formalism is not only a convenient representation of appropriate many-body formulations in the form of numerically feasible computational blocks, which plays a crucial role in the early stages of quantum computing development but also  provides a natural environment for capturing the sparsity characterizing correlated systems.

\subsection{Non-Hermitian CC Downfolding}
Active spaces play a central role in the development of CC  downfolding techniques and are defined by the subset $R$ of occupied active orbitals  ($R =\lbrace R_i, \; i=1,\ldots,x_R \rbrace$) and  a subset $S$ of active virtual orbitals ($S = \lbrace S_i, \; i=1,\ldots,y_s \rbrace$). Using many-body language, the excited Slater  determinants spanning the active space
(along with the reference function $|\Phi\rangle$) can be generated by generators 
$E^{a_l}_{i_l}=a_{a_l}^{\dagger} a_{i_l}$
($i_l\in R$ and $a_l\in S$) acting on the reference function $|\Phi\rangle$. These generators define the so-called $\mathfrak{g}^{(N)}(R,S)$ sub-algebra. Due to the utilization the particle-hole formalism all generators $E^{a_l}_{i_l}$ commute and the 
$\mathfrak{g}^{(N)}(R,S)$ is commutative. 
For the sake of the following analysis, it is convenient to  characterize various types of sub-algebras $\mathfrak{g}^{(N)}(R,S)$ by specifying the numbers of active occupied ($x_R$) and active virtual ($y_S$) orbitals, namely, 
$\mathfrak{g}^{(N)}(x_R,y_S)$.

As shown in  Refs.~\cite{safkk,downfolding2020t,kowalski2021dimensionality}, each sub-algebra  $\mathfrak{h} = \mathfrak{g}^{(N)}(R,S)$ induces partitioning of the cluster operator $T$:
\begin{equation}
T=T_{\rm int}(\mathfrak{h})+T_{\rm ext}(\mathfrak{h}) \;,
\label{deco1}
\end{equation}
where $T_{\rm int}(\mathfrak{h})$ belongs to $\mathfrak{h}$ while 
$T_{\rm ext}(\mathfrak{h})$ does no belong to $\mathfrak{h}$.
If the expansion $e^{T_{\rm int}(\mathfrak{h})}|\Phi\rangle$  produces all Slater determinants (of the same symmetry as the $|\Phi\rangle$ state) in the active space, we call $\mathfrak{h}$ the {\it sub-system embedding sub-algebra} for the CC formulation defined by the $T$ operator. 
In Ref.~\cite{safkk}, we showed that each standard CC approximation has its own class of SESs. 

The  existence of the SESs for standard CC approximations provides alternative ways for calculating CC energies, which can be obtained in contrast to the standard CC energy $E$ expression
\begin{equation}
  E=\langle\Phi|e^{-T}He^T|\Phi\rangle \;,
  \label{ccene}
\end{equation}
as an eigenvalue of the active-space non-Hermitian eigenproblem
\begin{equation}
  H^{\rm eff}(\mathfrak{h})
  e^{T_{\rm int}(\mathfrak{h})}|\Phi\rangle = 
  E e^{T_{\rm int}(\mathfrak{h})}|\Phi\rangle  \;.
\label{seseqh}
\end{equation}
where 
\begin{equation}
H^{\rm eff}(\mathfrak{h})=(P+Q_{\rm int}(\mathfrak{h})) \bar{H}_{\rm ext}(\mathfrak{h}) (P+Q_{\rm int}(\mathfrak{h}))\;
\label{heffses}
\end{equation}
and 
\begin{equation}
\bar{H}_{\rm ext}(\mathfrak{h})=e^{-T_{\rm ext}(\mathfrak{h})} H e^{T_{\rm ext}(\mathfrak{h})} \;.
\label{heffdef}
\end{equation}
The above result is known as the {\it SES-CC Theorem}.
In Eq.~(\ref{heffses}), $P$ stands for the projection operator onto reference function and  $Q_{\rm int}(\mathfrak{h})$
is a projection operator onto all excited Slater determinants  with respect to the reference function $|\Phi\rangle$ that correspond to $\mathfrak{h}$. 
One should also mention that the standard energy expression given by Eq.~(\ref{ccene}) can be reproduced from Eq.~(\ref{seseqh}) when $\mathfrak{h}$ represents the simplest case when there are no generators (both sets $R$ and $S$ are empty).
When SES-CC Theorem is applied to the exact cluster operator corresponding to the full coupled cluster approach, the lowest eigenvalues of effective Hamiltonians correspond to the exact ground-state full configuration interaction (FCI) energy.

Standard CC approximations (such as CCSD, CCSDT, CCSDTQ, etc. methods) are characterized by specific classes of SESs. For example, 
typical CCSD SESs are
$\mathfrak{g}^{(N)}(1_R,y_S)$ or $\mathfrak{g}^{(N)}(x_R,1_S)$ sub-algebras. For the CCSDTQ approach, corresponding SESs are of the 
$\mathfrak{g}^{(N)}(2_R,y_S)$ or $\mathfrak{g}^{(N)}(x_R,2_S)$ form. 
From these definitions, it is easy to see that lower-rank CC approximations are SESs for high-rank approaches. For example, the CCSD SESs 
$\mathfrak{g}^{(N)}(1_R,y_S)$  are SESs for the CCSDTQ formalism.

The SES-CC Theorem is flexible in the choice of active spaces, providing a number of alternative ways for calculating CC energies using effective Hamiltonians corresponding to various SESs. For example, for the CCSD approximation with a fixed molecular orbital basis, where the SES $\mathfrak{g}^{(N)}(R,S)$ defined at the orbital level\cite{sssc} contains either one occupied active orbital or one virtual active orbital, the number of different SESs $S_{\rm CCSD}$ and corresponding effective Hamiltonians that upon diagonalization reproduce the standard CCSD energy is 
\begin{equation}
S_{\rm CCSD} =n_o (2^{n_v}-1) + n_v (2^{n_o}-1) - n_o n_v \;.
\label{sccsd}
\end{equation}
This formula is a consequence of binomial expansion, in which $k$ active virtual/occupied orbitals can be chosen in 
$\binom{n_v}{k}$/$\binom{n_o}{k}$  different ways, where $n_o$ and $n_v$ stand for the numbers of correlated occupied and virtual orbitals, respectively.

The validity of the SES-CC Theorem has recently been confirmed numerically on the example of several benchmark systems. In addition to the standard spatial orbital-based definition of SES-generated active spaces, it was shown that the SES-CC Theorem also holds for the active spaces defined by a non-trivial number of active spin orbitals. In the extreme case, we demonstrated that the SES-CC Theorem is also satisfied 
for the active space describing one active electron "correlated" in two active $\alpha$-type spin-orbitals.\cite{sssc}

\subsection{Hermitian CC Downfolding}

The Hermitian form of the downfolded Hamiltonian is obtained as a consequence of utilizing DUCC representation of the wave function \cite{bauman2019downfolding,downfolding2020t}
\begin{equation}
      |\Psi\rangle=e^{\sigma_{\rm ext}(\mathfrak{h})} e^{\sigma_{\rm int}(\mathfrak{h})}|\Phi\rangle \;,
\label{ducc1}
\end{equation}
where $\sigma_{\rm ext}(\mathfrak{h})$ and $\sigma_{\rm int}(\mathfrak{h})$ are general-type anti-Hermitian operators
\begin{eqnarray}
\sigma_{\rm int}^{\dagger}(\mathfrak{h}) &=&  -\sigma_{\rm int}(\mathfrak{h}) \;, \label{sintah} \\
\sigma_{\rm ext}^{\dagger}(\mathfrak{h}) &=&  -\sigma_{\rm ext}(\mathfrak{h}) \;. \label{sintah2}
\end{eqnarray} 
In analogy to the non-Hermitian case, the $\sigma_{\rm int}(\mathfrak{h})$ operator is defined by parameters carrying only active spin-orbital labels and $\sigma_{\rm ext}(\mathfrak{h})$ operators are defined by parameters with at least one in-active spin-orbital label. 

The use of the  DUCC Ansatz (\ref{ducc1}), in analogy to the SES-CC case, leads to an alternative way of determining energy, which can be obtained by solving  active-space Hermitian eigenvalue problem:
\begin{equation}
      H^{\rm eff}(\mathfrak{h}) e^{\sigma_{\rm int}(\mathfrak{h})} |\Phi\rangle = E e^{\sigma_{\rm int}(\mathfrak{h})}|\Phi\rangle,
\label{duccstep2}
\end{equation}
where
\begin{equation}
      H^{\rm eff}(\mathfrak{h}) = (P+Q_{\rm int}(\mathfrak{h})) \bar{H}_{\rm ext}(\mathfrak{h}) (P+Q_{\rm int}(\mathfrak{h}))
\label{equivducc}
\end{equation}
and 
\begin{equation}
      \bar{H}_{\rm ext}(\mathfrak{h}) =e^{-\sigma_{\rm ext}(\mathfrak{h})}H e^{\sigma_{\rm ext}(\mathfrak{h})}\;.
\label{duccexth}
\end{equation}
When the external cluster amplitudes are known (or can be effectively approximated), the energy (or its approximation) can be calculated by diagonalizing the Hermitian effective/downfolded Hamiltonian (\ref{equivducc}) in the active space using various quantum or classical diagonalizers. 

For quantum computing applications second-quantized representation of $H^{\rm eff}(\mathfrak{h})$ is required. 
In the light of the non-commuting character of components defining the $\sigma_{\rm ext}(\mathfrak{h})$ operator, one has to rely on the finite-rank commutator expansions, i.e., 
 \begin{widetext}
\begin{equation}
H^{\rm eff}(\mathfrak{h}) \simeq
(P+Q_{\rm int}(\mathfrak{h}))(H +  \sum_{i=1}^{l}  \frac{1}{i!}[
 \ldots [H,\sigma_{\rm ext}(\mathfrak{h})],\ldots ],\sigma_{\rm ext}(\mathfrak{h})]_i (P+Q_{\rm int}(\mathfrak{h}))\;.
\end{equation}
 \end{widetext}
Due to the numerical costs associated with the contractions of multi-dimensional tensors and the rapidly expanding number of terms in this expansion, only approximations based on the inclusion of low-rank commutators are feasible. In recent studies, approximations based on single, double, and part of triple commutators were explored where one- and two-body interactions were retained in the second quantized form of $H^{\rm eff}(\mathfrak{h})$ were retained.
In practical applications, one also has to determine the approximate form of $\sigma_{\rm ext}(\mathfrak{h})$. For practical reasons, we used the following approximation
\begin{equation}
  \sigma_{\rm ext}(\mathfrak{h}) \simeq T_{\rm ext}(\mathfrak{h}) - T_{\rm ext}(\mathfrak{h})^{\dagger}\;,
  \label{sigext}
\end{equation}
where $T_{\rm ext}$ were defined through the external parts of the $T_1$ and $T_2$ operators obtained in CCSD calculations. 


\section{Quantum Flows}

\subsection{Non-Hermitian CC Flows}

In the case of non-Hermitian downfolding, 
the SES-CC Theorem can be used to form computational frameworks (quantum flow) that integrate eigenvalue problems \cite{kowalski2021dimensionality,bauman2022coupled}
\begin{equation}
  H^{\rm eff}(\mathfrak{h}_i)
  e^{T_{\rm int}(\mathfrak{h}_i)}|\Phi\rangle = 
  E e^{T_{\rm int}(\mathfrak{h}_i)}|\Phi\rangle  \;
  (i=1,\ldots,M_{\rm SES})\;,
\label{seseqhf}
\end{equation}
where $M_{\rm SES}$ is the total number of SESs or active space problems included in the flow. In Ref.~\cite{kowalski2021dimensionality,bauman2022coupled}, we demonstrated that a problem defined in this way is equivalent (at the solution) to the standard CC equations with cluster operator defined as 
a combination of all {\em unique} excitations included in 
$T_{\rm int}(\mathfrak{h}_i)$ $(i=1,\ldots,M_{\rm SES})$ operators, i.e., 
\begin{equation}
  T= \bigcup_{i=1}^{M} T_{\rm int}(\mathfrak{h}_i)
  \label{app3bb}
\end{equation}
and 
\begin{eqnarray}
Qe^{-T}He^T |\Phi\rangle = 0 \;, \label{cceqx1} \\
\langle\Phi|e^{-T}He^T |\Phi\rangle = E \;, \label{cceqx2}
\end{eqnarray}
where the $Q$ operator is a projection operator onto a sub-space of excited Slater determinants generated by the action of $T$ operator of  Eq.~(\ref{app3bb}) onto the reference function. The discussed equivalence  is known as the {\it Equivalence Theorem.}
Initially, as discussed in Ref.~\cite{safkk}, the quantum flows were introduced as a form of the invariance of the SES-CC Theorem upon separate rotations of occupied and virtual orbitals. 

Although the form given by Eqs.~(\ref{cceqx1}) and (\ref{cceqx2}) is generally better suited in canonical calculations to take advantage of parallel computing architectures, the representation given by Eq.~(\ref{seseqhf}) is well-poised to capture a general type of the sparsity characterizing  quantum systems.  This is because Eq.~(\ref{seseqhf}) represent reduced-dimensionality  
computational blocks representing quantum problems defined by non-Hermitian Hamiltonians $H^{\rm eff}(\mathfrak{h}_i)$. In analogy to the bi-variational CC formulations \cite{arponen83_311} one can introduce the left eigenvectors of active-space  Hamiltonians $H^{\rm eff}(\mathfrak{h}_i)$, using either CC-$\Lambda$ 
\begin{equation}
\langle\Phi|(1+\Lambda_{\rm int}(\mathfrak{h}_i)) \;\;
(i=1,\ldots,M_{\rm SES})\;,
\label{lambdah}
\end{equation}
or the extended CC formalism
\begin{equation}
\langle\Phi|e^{S_{\rm int}(\mathfrak{h}_i)} \;\;
(i=1,\ldots,M_{\rm SES})\;,
\label{extcc}
\end{equation}
where $\Lambda_{\rm int}(\mathfrak{h}_i)$ and $S_{\rm int}(\mathfrak{h}_i)$ are de-excitation operators acting in the corresponding active spaces,
to form one-particle reduced density matrices ${\bf \gamma}(\mathfrak{h}_i)$. For the $\Lambda$-CC formalism
the matrix elements of the ${\bf \gamma}(\mathfrak{h}_i)$ are given by the formula 
\begin{eqnarray}
\gamma^{p}_{q}(\mathfrak{h}_i) &=& 
\langle\Phi|(1+\Lambda_{\rm int}(\mathfrak{h}_i))a_p^{\dagger}a_q
e^{T_{\rm int}(\mathfrak{h}_i)}|\Phi\rangle \;,\; \\
&& a_p^{\dagger} a_q \in \mathfrak{h_i} \;,\;
\forall_{i=1,\ldots,M_{\rm SES}}\;.
\label{ttzz}
\end{eqnarray}
The above construct, in contrast to the existing local CC formulations where the one-particle reduced density matrices are postulated, allows one to introduce them in a natural way.
A more detailed analysis of the local formulations stemming from the Equivalence Theorem is discussed in Refs.~\cite{kowalski2021dimensionality,bauman2022coupled}.
This procedure can also be extended to systems driven by
different types of interactions, such as in nuclear structure
theory or quantum lattice models, where the extension of
the standard local CC formulations as used in quantum
chemistry may be less obvious.

\subsection{Hermitian CC Flows}

Using non-Hermitian formulation as a guide, the idea of quantum flow can be generalized to the DUCC case. 
We start our analysis by assuming that we would like to perform DUCC effective simulations for SES $\mathfrak{h}$ problem expressed in Eq.~(\ref{duccstep2}) for an active space that is too big to be handled either in classical or quantum computing. 
We will assume that external amplitudes $\sigma_{\rm ext}(\mathfrak{h})$
can be effectively approximated. For simplicity we will introduce a new DUCC Hermitian Hamiltonian $A(\mathfrak{h})$ which is defined 
as $H^{\rm eff}(\mathfrak{h})$ or its approximation in the 
$(P+Q_{\rm int}(\mathfrak{h}))$ space (in the simplest case it can be just 
the $(P+Q_{\rm int}(\mathfrak{h}))H(P+Q_{\rm int}(\mathfrak{h}))$ operator). 
We will denote $A(\mathfrak{h})$ simply by $A$.
Next, we assume that    excitations  in  $\mathfrak{h}$ that are relevant 
to the state of interest can be captured by excitation sub-algebras: $\mathfrak{h}_1$, $\mathfrak{h}_2$, $\ldots$, 
$\mathfrak{h}_M$, where, in analogy to the SR-CC case, we admit the possibility of "sharing" excitations/de-excitations between these sub-algebras. We also assume that the number of excitations belonging to  each 
$\mathfrak{h}_i$ $(i=1,\ldots,M)$ is significantly smaller  than the number of excitations in $\mathfrak{h}$ and therefore 
numerically tractable in simulations. 

The  $A(\mathfrak{h})$ Hamiltonian and the
$(P+Q_{\rm int}(\mathfrak{h}))$ space can be treated as a starting point for the secondary DUCC  decompositions
generated by sub-system algebras
$\mathfrak{h}_i$ $(i=1,\ldots,M)$
defined above,
i.e.,
\begin{equation}
  A^{\rm eff}(\mathfrak{h}_i)
  e^{\sigma_{\rm int}(\mathfrak{h}_i)}|\Phi\rangle = 
  E e^{\sigma_{\rm int}(\mathfrak{h}_i)}|\Phi\rangle \;
 \;(i=1,\ldots,M)\;
\label{seseqh22}
\end{equation}
or in the VQE-type variational representation as 
\begin{equation}
\min_{{\bm \theta}(\mathfrak{h}_i)}   
\langle\Psi({\bm \theta}(\mathfrak{h}_i))|A^{\rm eff}(\mathfrak{h}_i)|
\Psi({\bm \theta}(\mathfrak{h}_i))\rangle \;\;
(i=1,\ldots,M)\;,
\label{vqeduccx1}
\end{equation}
where $|
\Psi({\bm \theta}(\mathfrak{h}_i))\rangle$ approximates 
$e^{\sigma_{\rm int}(\mathfrak{h}_i)}|\Phi\rangle$.
Each $A^{\rm eff}(\mathfrak{h}_i)$ is defined as
\begin{equation}
      A^{\rm eff}(\mathfrak{h}_i) = (P+Q_{\rm int}(\mathfrak{h}_i)) \bar{A}_{\rm ext}(\mathfrak{h}_i) (P+Q_{\rm int}(\mathfrak{h}_i))
\label{equivducc66}
\end{equation}
and 
\begin{equation}
      \bar{A}_{\rm ext}(\mathfrak{h}_i) =e^{-\sigma_{\rm ext}(\mathfrak{h}_i)}A  e^{\sigma_{\rm ext}(\mathfrak{h}_i)},
\label{duccexth89}
\end{equation}
where we defined external $\sigma_{\rm ext}(\mathfrak{h}_i)$ operator
with respect to $\mathfrak{h}$ or
$(P+Q_{\rm int}(\mathfrak{h}))$ space (i.e. cluster amplitudes defining $\sigma_{\rm ext}(\mathfrak{h}_i)$ must carry at last one index belonging to active spin orbitals defining $\mathfrak{h}$ and not belonging to the
set of active spin orbtitals defining $\mathfrak{h}_i$).
In other words, sub-algebras $\mathfrak{h}_i$  generate active sub-spaces in the larger 
active space $\mathfrak{h}$, i.e.,
$(P+Q_{\rm int}(\mathfrak{h}_i)) \subset (P+Q_{\rm int}(\mathfrak{h}))$. 
Due to the non-commutativity of components defining $\sigma$-operators, connecting  DUCC computational blocks given by Eq.~(\ref{seseqh22}) or Eq.~(\ref{vqeduccx1}) directly into a flow is a rather challenging task. To address these issues (see Ref.~\cite{kowalski2021dimensionality}) and define practical DUCC flow, we will discuss the algorithm that combines secondary downfolding steps with Trotterization of the unitary CC operators. 
Let us  assume that the $\sigma_{\rm int}(\mathfrak{h})$ operator can be approximated by $\sigma_{\rm int}(\mathfrak{h}_i)
(i=1,\ldots,M)$, i.e.,
\begin{equation}
  \sigma_{\rm int}(\mathfrak{h}) \simeq
  \sum_{i=1}^M \sigma_{\rm int}(\mathfrak{h}_i) +
  X(\mathfrak{h},\mathfrak{h}_1,\ldots,\mathfrak{h}_M)\;,
  \label{trott1}
\end{equation}
where the  $X(\mathfrak{h},\mathfrak{h}_1,\ldots,\mathfrak{h}_M)$ operator (or $X$ for short) eliminates possible overcounting of the "shared" amplitudes. It enables to  re-express $\sigma_{\rm int}(\mathfrak{h})$ as
\begin{equation}
\sigma_{\rm int}(\mathfrak{h}) = \sigma_{\rm int}(\mathfrak{h}_i) + R(\mathfrak{h}_i) \;\;
(i=1,\ldots,M)\;,
\label{sisplit}
\end{equation}
where 
\begin{equation}
R(\mathfrak{h}_i) = 
^{(i)}\sum_{j=1}^M \; \sigma_{\rm int}(\mathfrak{h}_j) +
  X
\end{equation}
and $^{(i)}\sum_{j=1}^M$ designates the sum where the $i$-th element is neglected.
Consequently, we get 
\begin{equation}
  e^{\sigma_{\rm int}(\mathfrak{h})}|\Phi\rangle=
  e^{\sigma_{\rm int}(\mathfrak{h}_i)+R(\mathfrak{h}_i)}|\Phi\rangle \;\;(i=1,\ldots,M)\;.
  \label{ffqq1}
\end{equation}
Using the Trotter formula, we can approximate the right-hand side of Eq.~(\ref{ffqq1}) for a given $j$  as 
\begin{equation}
   e^{\sigma_{\rm int}(\mathfrak{h})}|\Phi\rangle \simeq
   ( e^{R(\mathfrak{h}_i)/N}
   e^{\sigma_{\rm int}(\mathfrak{h}_i)/N})^N |\Phi\rangle \;.
\end{equation}
Introducing auxiliary operator $G^{(N)}_i$
\begin{equation}
  G^{(N)}_i=( e^{R(\mathfrak{h}_i)/N}
   e^{\sigma_{\rm int}(\mathfrak{h}_i)/N})^{N-1} e^{R(\mathfrak{h}_i)/N}
   \;\;(i=1,\ldots,M)\;,
\end{equation}
the "internal" wave function (\ref{ffqq1}) can be expressed as 
\begin{equation}
  e^{\sigma_{\rm int}(\mathfrak{h})}|\Phi\rangle \simeq G^{(N)}_i e^{\sigma_{\rm int}(\mathfrak{h}_i)/N} |\Phi\rangle
  \;\;(i=1,\ldots,M)\;,
  \label{ubi1}
\end{equation}
where  $G^{(N)}_i$ is a complicated function of all $\sigma_{\rm int}(\mathfrak{h}_j)\; 
(j=1,\ldots,M)$ and the above expression does not decouple $\sigma_{\rm int}(\mathfrak{h}_i)$ from the $G^{(N)}_i$ term. 
However, using this  expression, one can define the practical way of determining computational blocks for flow equations. 
To this end, let us introduce the expansion in Eq.~(\ref{ubi1}) to Eq.~(\ref{duccstep2}) (with $H^{\rm eff}(\mathfrak{h})$ replaced by the $A$ operator), pre-multiply both sides by $\lbrack G^{(N)}_i \rbrack^{-1}$, and project onto $(P+Q_{\rm int}(\mathfrak{h}_i))$ sub-space, which leads to non-linear eigenvalue problems
\begin{widetext}
\begin{equation}
(P+Q_{\rm int}(\mathfrak{h}_i)) \lbrack G^{(N)}_i\rbrack ^{-1} A G^{(N)}_i e^{\sigma_{\rm int}(\mathfrak{h}_i)/N} |\Phi\rangle \simeq E e^{\sigma_{\rm int}(\mathfrak{h})_i)/N} |\Phi\rangle
\; (i=1,\ldots,M)
\;.
\label{upr1}
\end{equation}
\end{widetext}
These equations define computational blocks for the  DUCC flow. 
To make practical use of Eqs. (\ref{upr1}) let us linearize them by defining the downfolded Hamiltonian $\Gamma_i^{(N)}$,
$\Gamma_i^{(N)}= (P+Q_{\rm int}(\mathfrak{h}_i)) \lbrack G^{(N)}_i\rbrack ^{-1} A G^{(N)}_i  (P+Q_{\rm int}(\mathfrak{h}_i))$ as a function of all
$\sigma_{\rm int}(\mathfrak{h}_j) \; (j=1,\ldots,M)$ from the previous flow cycle(s) ($pc$). We will symbolically designate this fact by using special notation for $\Gamma_i^{(N)}$ effective Hamiltonian, i.e., 
$\Gamma_{i}^{(N)}(pc)$ Hamiltonian. 
Now, we replace eigenvalue problems in Eq.~(\ref{upr1}) by an optimization procedures described by  Eq.~(\ref{vqeduccx1}) which also offer an easy way to 
deal with "shared" amplitudes.  Namely, if, in analogy to SR-CC flow, we establish an ordering of
$\mathfrak{h}_i$ sub-algebras, with $\mathfrak{h}_1$ corresponding to the importance of active spaces with respect to the  wave function of interest. Then 
in the $\mathfrak{h}_i$ problem we partition  set of parameters 
${\bm \theta}_N(\mathfrak{h}_i)$ into subset ${\bm \theta}_N^{\rm CP}(\mathfrak{h}_i)$ that refers to  common pool of  
amplitudes determined in preceding steps (say, for $\mathfrak{h}_j \; (j=1,\ldots,i-1)$) and subset  ${\bm \theta}_N^{\rm X}(\mathfrak{h}_i)$ that is uniquely determined in the $\mathfrak{h}_i$ minimization step, i.e, 
\begin{widetext}
\begin{equation}
\min_{{\bm \theta}_N^{\rm X}(\mathfrak{h}_i)}   
\langle\Psi({\bm \theta}_N^{\rm X}(\mathfrak{h}_i),{\bm \theta}_N^{\rm CP}(\mathfrak{h}_i))
|
\Gamma_{i}^{(N)}(pc)
|
\Psi({\bm \theta}_N^{\rm X}(\mathfrak{h}_i),{\bm \theta}_N^{\rm CP}(\mathfrak{h}_i))\rangle 
\;\;(i=1,\ldots,M)\;,
\label{vqetrott}
\end{equation}
\end{widetext}
where $|\Psi({\bm \theta}_N^{\rm X}(\mathfrak{h}_i),{\bm \theta}_N^{\rm CP}(\mathfrak{h}_i))\rangle$ approximates 
$e^{\sigma_{\rm int}(\mathfrak{h}_i)/N} |\Phi\rangle$.
In this way, each computational block coupled into a flow corresponds to a minimization procedure that optimizes parameters
${\bm \theta}_N^{\rm X}(\mathfrak{h}_i)$ using quantum algorithms such as the VQE approach. 
At the end of the iterative cycle, once all amplitudes are converged, in contrast to the SR-CC flows, the energy is calculated using  $\mathfrak{h}_1$ problem
as an expectation value of the $\Gamma_{1}^{(N)}$ operator. 
The discussed formalism introduces a broad class of control parameters defining each computational step's dimensionality. 
These are the numbers of occupied/unoccupied active orbitals defining $\mathfrak{h}_i$ sub-algebras $x_{R_i}$/$y_{S_i}$, respectively. 

An essential  feature of the DUCC flow equation is associated with the fact that each computational block (\ref{vqetrott})  can be encoded using a much smaller number of qubits compared to the  qubits requirement associated with the original problem. 
This observation significantly simplifies the qubit  encoding of the effective Hamiltonians included in  quantum DUCC flows, especially in formulations based on the utilization of localized molecular basis set (for quantum algorithms exploiting locality of interactions, see Refs.~\cite{mcclean2014exploiting,otten2022localized}).

\section{Time-Dependent CC Extensions}

The  SES-CC-based downfolding techniques could also be extended to the  time-dependent domain.\cite{downfolding2020t,bauman2022coupled} 
As in the stationary case,  we will assume a general partitioning of the 
time-dependent cluster operator $T(t)$ into its
internal ($T_{\rm int}(\mathfrak{h},t)$) and external ($T_{\rm ext}(\mathfrak{h},t)$) parts (we also assume that the employed molecular orbitals are time-independent), i.e,
\begin{equation}
|\Psi(t)\rangle 
= e^{T_{\rm ext}(\mathfrak{h},t)} e^{T_{\rm int}(\mathfrak{h}, t)} |\Phi\rangle \;,
\forall \mathfrak{h} \in SES  \;. \label{step2} 
\end{equation}
For generality, we also include phase factor $T_0(\mathfrak{h},t)$ in the definition of the  $T_{\rm int}(\mathfrak{h},t)$ operator.
After substituting (\ref{step2}) into time-dependent  Schr\"odinger equation and utilizing properties of SES algebras, we demonstrated that the ket-dynamics of the sub-system wave function 
$e^{T_{\rm int}(\mathfrak{h},t)}|\Phi\rangle$
corresponding to arbitrary SES $\mathfrak{h}$
\begin{equation}
i\hbar \frac{\partial }{\partial t} e^{T_{\rm int}(\mathfrak{h},t)} |\Phi\rangle = H^{\rm eff}(\mathfrak{h},t)  e^{T_{\rm int}(\mathfrak{h},t)} |\Phi\rangle \;,
\label{cool2}
\end{equation}
where 
\begin{equation}
H^{\rm eff}(\mathfrak{h},t) = (P+Q_{\rm int}(\mathfrak{h})) \bar{H}_{\rm ext}(\mathfrak{h},t) (P+Q_{\rm int}(\mathfrak{h})) \;
\label{lemma2}
\end{equation}
and 
\begin{equation}
\bar{H}_{\rm ext}(\mathfrak{h},t) = e^{-T_{\rm ext}(\mathfrak{h},t)} H e^{T_{\rm ext}(\mathfrak{h},t)} \;.
\label{hbart}
\end{equation}
If $T_{\rm ext}(\mathfrak{h},t)$ operator is known or can be efficiently approximated, then the dynamics of the entire system can be described by effective Hamiltonian $H^{\rm eff}(\mathfrak{h},t)$.

In analogy to the stationary cases, various sub-systems computational blocks can be integrated into a flow enabling sampling of large sub-spaces of Hilbert space through  
through a number of coupled reduced-dimensionality  problems (time-dependent quantum flows), i.e.,
\begin{equation}
i\hbar \frac{\partial }{\partial t} e^{T_{\rm int}(\mathfrak{h}_i,t)} |\Phi\rangle = H^{\rm eff}(\mathfrak{h}_i,t)  e^{T_{\rm int}(\mathfrak{h}_i,t)} |\Phi\rangle \;,\;
(i=1,\ldots,M_{\rm SES})\;.
\label{seseqht}
\end{equation}
Given analogies between stationary CC flow equations based on the localized orbitals and local CC formulations developed in the last few decades in quantum chemistry, 
time-dependent flow equations given by Eq.~(\ref{seseqht})
can be utilized to  design  reduced-scaling variants of the time-dependent CC formulations.

The time-dependent variant of the DUCC Ansatz is 
represented by  the normalized time-dependent wave function 
$|\Psi_{\rm DUCC}(\mathfrak{h},t)\rangle$,
\begin{equation}
|\Psi_{\rm DUCC}(\mathfrak{h},t)\rangle = e^{\sigma_{\rm ext}(\mathfrak{h},t)} e^{\sigma_{\rm int}(\mathfrak{h},t)} |\Phi\rangle \;,
\forall \mathfrak{h} \in SES \;,
\label{ducct}
\end{equation}
where $\sigma_{\rm int}(\mathfrak{h},t)$ and $\sigma_{\rm ext}(\mathfrak{h},t)$ are 
general-type time-dependent anti-Hermitian operators 
\begin{eqnarray}
\sigma_{\rm int}(\mathfrak{h},t)^{\dagger} &=& - \sigma_{\rm int}(\mathfrak{h},t)\;,  \label{ucc1} \\
\sigma_{\rm ext}(\mathfrak{h},t)^{\dagger} &=& - \sigma_{\rm ext}(\mathfrak{h},t)\;.  \label{ucc2}
\end{eqnarray}
Again, as in the SES-CC case, the dynamics of the entire system are given by the active-space  time-dependent effective Hamiltonian
$H^{\rm eff}(\sigma_{\rm ext}(\mathfrak{h},t),\frac{\partial \sigma_{\rm ext}(\mathfrak{h},t)}{\partial t})$
\begin{widetext}
\begin{equation}
H^{\rm eff}(\sigma_{\rm ext}(\mathfrak{h},t),\frac{\partial \sigma_{\rm ext}(\mathfrak{h},t)}{\partial t}) = 
(P+Q_{\rm int} )\lbrace  \bar{H}_{\rm ext}(\mathfrak{h},t) -i\hbar 
A(\sigma_{\rm ext}(\mathfrak{h},t), \frac{\partial \sigma_{\rm ext}(\mathfrak{h},t)}{\partial t})  \rbrace
(P+Q_{\rm int})  \;.
\label{gammae} 
\end{equation}
\end{widetext}
where anti-Hermitian operator $A(\sigma_{\rm ext}(\mathfrak{h},t), \frac{\partial \sigma_{\rm ext}(\mathfrak{h},t)}{\partial t})$ is expressed as 
\begin{equation}
A(\sigma_{\rm ext}(\mathfrak{h},t), \frac{\partial \sigma_{\rm ext}(\mathfrak{h},t)}{\partial t}) = 
\sum_{k=0}^{\infty} \frac{(-1)^k}{(k+1)!} I_k(\sigma_{\rm ext}(\mathfrak{h},t), \frac{\partial \sigma_{\rm ext}(\mathfrak{h},t)}{\partial t})
\label{asigt}
\end{equation}
and
\begin{widetext}
\begin{equation}
I_k(X(\mathfrak{h},t),\frac{\partial X(\mathfrak{h},t)}{\partial t}) = \underbrace{[X(\mathfrak{h},t),[X(\mathfrak{h},t), \ldots  [X(\mathfrak{h},t),[X(\mathfrak{h},t),\frac{\partial X(\mathfrak{h},t)}{\partial t}]]\ldots ]]}_{k \; {\rm times}}\;.
\label{ik}
\end{equation}
\end{widetext}
If the fast-varying in time part of the wave function (or $\sigma_{\rm ext}(\mathfrak{h},t)$-dependent part of the wave function) is known or 
can be efficiently approximated, then the slow-varying  dynamic (captured by the proper  choice of the active space and $\sigma_{\rm int}(\mathfrak{h},t)$ operator)
of the entire system can be described as a sub-system dynamics generated by the Hermitian
$H^{\rm eff}(\sigma_{\rm ext}(\mathfrak{h},t), \frac{\partial \sigma_{\rm ext}(\mathfrak{h},t)}{\partial t})$ operator.  This decoupling of various time regimes (slow- vs. fast-varying components)  
is analogous to decoupling high- and low-energy Fermionic degrees of freedom in stationary formulations of the SES-CC and DUCC formalisms.

\section{Green's Function Applications}

The CC Green's function formulations 
have recently evolved into important formulations to describe spectral functions in various energy regimes \cite{nooijen92_55,nooijen93_15,meissner93_67,schirmer2004intermediate,mertins96_2140,kkgfcc2,kkgfcc3} and as high-accuracy solvers for quantum embedding formulations.
Following original formulations
based on the CC bi-variational approach, the corresponding frequency-dependent Green's function for an $N$-particle system can be expressed as
\begin{eqnarray}
&&G_{pq}(\omega) =  \nonumber \\
&&\langle\Phi|(1+\Lambda)e^{-T} a_q^{\dagger} (\omega+(H-E)- \text{i} \eta)^{-1} a_p e^T |\Phi\rangle + \nonumber \\
&& \langle\Phi|(1+\Lambda)e^{-T} a_p (\omega-(H-E)+ \text{i} \eta)^{-1} a_q^{\dagger} e^T |\Phi\rangle \;,
\label{gfxn0}
\end{eqnarray}
where $\omega$ denotes the frequency parameter, and the imaginary part $\eta$ is often called a broadening factor. The cluster operator $T$ and de-excitation operator $\Lambda$ define  correlated ket ($|\Psi\rangle$) and bra 
($\langle\Psi|$)
ground-state wave functions
for $N$-electron system 
\begin{eqnarray}
|\Psi\rangle&=&e^T |\Phi\rangle \;,
\label{ccfun} \\
\langle\Psi| &=& \langle\Phi|(1+\Lambda)e^{-T} \;.
\label{ccfunl}
\end{eqnarray}
The ground-state energy $E_0$, and the amplitudes defining $T$ and $\Lambda$ operators are obtained from the following sequence of CC equations.

To combine GFCC and DUCC formalisms we replace $T$, $\Lambda$, and $H$ operators in Eq.~(\ref{gfxn0}) by  cluster ($\tilde{T}_{\rm int}$), de-excitation ($\tilde{\Lambda}_{\rm int}$),  and Hermitian  $H^{\rm eff}(\mathfrak{h})$ [Eq.~(\ref{equivducc}), which is further denoted as $\Gamma$]  operators acting in the some active space generated by $\mathfrak{h}$ (for the notational simplicity we also skip the $\mathfrak{h}$ symbol).\cite{baumanpengGFDUCC}
We will also consider the case when 
the set of active orbitals consists of all occupied orbitals and a small subset of active virtual orbitals (containing $n_{v}^{\rm act}$ active virtual orbitals), where, in general, $n_{v}^{\rm act}\ll n_v$, where $n_v$ designates the total number of virtual orbitals.
The standard CC equations for $T$, CC energy $E$, and $\Lambda$ are 
are replaced by their ``active" counterparts
\begin{widetext}
\begin{eqnarray}
Q_{\rm int} e^{-\widetilde{T}_{\rm int}}\Gamma 
e^{\widetilde{T}_{\rm int}}|\Phi\rangle &=& 0 ~, \label{eq:cceqa} \\
\langle\Phi|e^{-\widetilde{T}_{\rm int}}\Gamma 
e^{\widetilde{T}_{\rm int}}|\Phi\rangle &=& E_0^{\rm int} ~, \label{eeqa} \\
\langle\Phi|(1+\widetilde{\Lambda}_{\rm int}) e^{-\widetilde{T}_{\rm int}}\Gamma e^{\widetilde{T}_{\rm int}} Q_{\rm int} &=& E_0^{\rm int} \langle \Phi|(1+\widetilde{\Lambda}_{\rm int})Q_{\rm int} ~. \label{eqla} 
\end{eqnarray}
\end{widetext}
The coupled cluster Green's function employing  the DUCC Hamiltonian $\Gamma$ can be expressed for active orbitals as follows
 \begin{widetext}
\begin{eqnarray}
G_{PQ}^{\rm DUCC}(\omega) &=& \langle\Phi|(1+\widetilde{\Lambda}_{\rm int})e^{-\widetilde{T}_{\rm int}} a_Q^{\dagger} (\omega+(\Gamma-E_0^{\rm int})- \text{i} \eta)^{-1} a_P e^{\widetilde{T}_{\rm int}} |\Phi\rangle + \notag \\
&& \langle\Phi|(1+\widetilde{\Lambda}_{\rm int})e^{-\widetilde{T}_{\rm int}} a_P (\omega-(\Gamma-E_0^{\rm int})+ \text{i} \eta)^{-1} a_Q^{\dagger} e^{\widetilde{T}_{\rm int}} |\Phi\rangle \;, 
\label{gfxn0ac}
\end{eqnarray}
\end{widetext}
where indices $P,Q,\ldots$ designate active spin orbitals. 
Again, applying the resolution of identity $e^{-\widetilde{T}_{\rm int}}e^{\widetilde{T}_{\rm int}}$ in 
the above equation, one gets the following expressions for DUCC Green's function matrix elements
\begin{widetext}
\begin{eqnarray}
G_{PQ}^{\rm DUCC}(\omega) = 
&&\langle\Phi|(1+\widetilde{\Lambda}_{\rm int}) \overline{a_Q^{\dagger}}^{\rm int} (\omega+\overline{\Gamma}_N- \text{i} \eta)^{-1} 
\overline{a_P}^{\rm int} |\Phi\rangle + \notag \\
&& \langle\Phi|(1+\widetilde{\Lambda}_{\rm int}) \overline{a_P}^{\rm int} (\omega-\overline{\Gamma}_N+ \text{i} \eta)^{-1} 
\overline{a_Q^{\dagger}}^{\rm int} |\Phi\rangle \;,
\label{gfxn1ints}
\end{eqnarray}
\end{widetext}
where we used the following definitions:
\begin{eqnarray}
\overline{\Gamma} &=& e^{-\widetilde{T}_{\rm int}} \Gamma ~e^{\widetilde{T}_{\rm int}} \;, \label{gbarac} \\
\overline{\Gamma}_N &=& \overline{\Gamma}-E_0^{\rm int}  \;,\label{gnomac}\\
\overline{a_P}^{\rm int} &=& e^{-\widetilde{T}_{\rm int}} a_P ~e^{\widetilde{T}_{\rm int}}, \label{apac}\\
\overline{a_Q^\dagger}^{\rm int} &=& e^{-\widetilde{T}_{\rm int}} a_Q^\dagger ~e^{\widetilde{T}_{\rm int}}. \label{aqac} 
\end{eqnarray}
In the  active-space driven DUCC-GFCC approach, the $X_p(\omega)$ and $Y_q(\omega)$
operators are replaced by $X_P^{\rm int}(\omega)$ and $Y_Q^{\rm int}(\omega)$, respectively, which are given by the following expressions:
\begin{eqnarray}
X_P^{\rm int}(\omega) &=& \sum_{I} x^I(P, \omega)^{\rm int}  a_I  + \sum_{I<J,A} x^{IJ}_A(P, \omega)^{\rm int} a_A^{\dagger} a_J a_I +\ldots ~, \label{xpac}  \\
Y_Q^{\rm int}(\omega) &=& \sum_{A} y_A(Q, \omega)^{\rm int} a_A^\dagger  + \sum_{I,A<B} y^{I}_{AB}(Q, \omega)^{\rm int} a_A^{\dagger} a_B^\dagger a_I +\ldots ~,  
\label{yqacp}
\end{eqnarray}
where indices $I, J,\ldots$ and $A, B,\ldots$ refer to active occupied and unoccupied spin orbitals indices, respectively (again, in the present discussion, we assume that all occupied spin orbitals are treated as active). These operators  satisfy
\begin{eqnarray}
(\omega+\overline{\Gamma}_N - \text{i} \eta )X_P^{\rm int}(\omega)|\Phi\rangle = 
\overline{a_P}^{\rm int} |\Phi\rangle \;, \label{eq:xpac} \\
(\omega-\overline{\Gamma}_N + \text{i} \eta )Y_Q^{\rm int}(\omega)|\Phi\rangle = 
\overline{a_Q^\dagger}^{\rm int} |\Phi\rangle \;,\label{eq:yqac}
\end{eqnarray}
and the $G_{PQ}^{\rm DUCC}(\omega$) is given by the expression
\begin{eqnarray}
G_{PQ}^{\rm DUCC}(\omega) = 
&&\langle\Phi|(1+\Lambda_{\rm int}) \overline{a_Q^{\dagger}}^{\rm int}  X_P^{\rm int}(\omega) |\Phi\rangle + \notag \\
&& \langle\Phi|(1+\Lambda_{\rm int}) 
\overline{a_P}^{\rm int} Y_Q^{\rm int}(\omega) |\Phi\rangle \;.
\label{gfxn2}
\end{eqnarray}
We demonstrated that the combined GFCC and DUCC frameworks  reproduce the main features of the standard  GFCCSD spectral function. In a series of test calculations, we demonstrated that increasing active space size leads to monotonic improvements in the location of  peaks obtained with the  DUCC-GFCCSD approach with respect to the full GFCCSD results. We attribute this behavior to the presence of dynamical (out-of-active-space) correlation effects encapsulated in each of  DUCC effective Hamiltonians. In contrast to the DUCC-GFCCSD formalism, the utilization of active space bare Hamiltonians leads to less consistent results for the peak positions.
The utilization of the DUCC effective Hamiltonians can also significantly reduce the cost of the GFCC calculations for the energy regime embraced by the corresponding active space.

\section{Review of Applications}

This section briefly reviews several exemplary application areas and numerical studies involving various CC downfolding formalisms. 

\subsection{Numerical Validation of the SES-CC Theorem}
The SES-CC Theorem has recently been validated on the example of several benchmark systems (H4, H6, H8 models) used to test CC methodologies in situations corresponding to the presence of weak and strong correlation effects (see Ref.~\cite{sssc}). 
We numerically verified the SES-CC Theorem using various active spaces corresponding to physically meaningful active spaces, capturing the most important correlation for the ground-state wave function description in the valence region as well as for the active spaces that are remotely related to the ground-state correlation effects. To this end, we used two approaches, CCSD and CCSDTQ, to calculate the eigenvalues of effective/downfolded Hamiltonians. In all cases considered in Ref.~\cite{sssc}, we were able to reproduce the CCSD or CCSDTQ energies, obtained  using the standard expression for the energy, as lowest-energy eigenvalues of the effective  Hamiltonian. In the extreme case, we used a spin-orbital-based  definition of active to correlate a single electron, which also resulted in reproducing standard CC energies.

\subsection{Approximations Based on  Quantum Flows}
In Ref.~\cite{safkk}, we introduced QFs based on the 
ordered flow of the 
$\mathfrak{g}^{(N)}(2_R)$ sub-algebras: 
\begin{equation}
\mathfrak{g}^{(N)}(2_{R_1})
\stackrel{\mathrm{\rm passing T}}{\longrightarrow}
\mathfrak{g}^{(N)}(2_{R_2}) 
\stackrel{\mathrm{\rm passing T}}{\longrightarrow}
\ldots
\stackrel{\mathrm{\rm passing T}}{\longrightarrow}
\mathfrak{g}^{(N)}(2_{R_{\rm final}})
\label{flow2}
\end{equation}
where SESs are ordered according to some importance criterium, for example, corresponding to  a descending order with respect to the sum of orbital energies 
($\epsilon_{k_i}+\epsilon_{l_i}$) corresponding to orbitals included in  $R_i$ sets. 
The Equivalence Theorem states that such a flow (in the discussed case, all $\mathfrak{g}^{(N)}(2_R)$-generated active space problems are integrated) is equivalent to  
the CC formalism defined by the following cluster operator $T$:
\begin{equation}
T\simeq T_1+T_2 + \sum_{R}  T_{\rm int,3}(\mathfrak{g}^{(N)}(2_{R}))
+\sum_{R}  T_{\rm int,4}(\mathfrak{g}^{(N)}(2_{R})) \;.
\label{scnf}
\end{equation}
This approach is further referred to as the self-consistent 
sub-algebra  flow CC method (SCSAF-CC).
In contrast to  the class of the so-called active-space CC approaches (see Refs.~\cite{piecuchactive} and references therein), this formulation includes classes of triply and quadruply excitations (in addition to all possible single and double excitations)  corresponding to triply and quadruply excited amplitudes corresponding to active-space problems included in the flow. 

The performance of the SCSAF-CC methods was evaluated on the examples involving single and double bond-breaking processes. For example, for the F$_2$ benchmark system, the non-parallel error (NPE)  of the SCSAF-CC formalism given by Eq.~(\ref{scnf}) in describing ground-state potential energy surface
is comparable to the NPSs yielded by the 4-reference reduced multi-reference CCSD(T) approach (see Ref.~\cite{safkk} for more details). As demonstrated in Ref.~\cite{safkk}, the SCSAF-CC formalism provided an efficient way for perturbative corrections due to  triple or quadruple excitations not included in the iterative SCSAF-CC formulation given by Eq.~\ref{scnf}). It was shown that perturbative SCSAD-CC methods could bypass typical problems with perturbative inclusion of higher-rank clusters in situations with strong correlation effects. 

Similarly to the ground-state case, the SCSAF-CC methods can be extended to their equation-of-motion CC (EOMCC) \cite{bartlett89_57,bartlett93_414,stanton93_7029} formulations using a similar manifold of excitations as in the ground-state case. For example, the state-specific excitation operator for $K$-th state $X(K)$ corresponding to the $\mathfrak{g}^{(N)}(2_R)$ flow as in Eq.~(\ref{scnf}) is given by the expansion:
\begin{eqnarray}
X(K) &\simeq& X(K)_0 + X(K)_1+X(K)_2 + \sum_{R}  X(K)_{\rm int,3}(\mathfrak{g}^{(N)}(2_{R})) \nonumber \\
&&+\sum_{R}  X(K)_{\rm int,4}(\mathfrak{g}^{(N)}(2_{R})) \;.
\label{xscnf}
\end{eqnarray}
These EOMCC-type extensions have been shown to properly capture excited-state correlation effects for singly excited states and a more challenging class of excited states dominated by double excitations.

\subsection{Quantum Computing}
The Hermitian CC downfolding plays a vital role in realizing quantum computing applications in computational chemistry with limited resources defined by  Noisy Intermediate-State Quantum (NISQ) devices. In particular, a significant effort has been expanded to provide frameworks that significantly reduce the size of the virtual space. Using these techniques, we could adequately reproduce total ground-state energies for systems described by 50-70 molecular basis functions employing small-size active spaces (5-15 molecular orbitals) and various types of solvers, including VQE and QPE quantum algorithms. One should stress that quantum simulations for systems described by 50-70 orbitals are currently beyond reach, which is a net effect of the required logical qubits, quantum errors, quantum circuit depth, and large numbers of fermionic degrees of freedom (amplitudes) to be included to achieve the necessary level of accuracy. The last factor translates into a massive number of quantum measurements concerning VQE class of methods. 

Special consideration is required when constructing approximate downfolded Hamiltonians is required to enable downfolding methods in quantum computing. The key factors that need to be taken care of when approximating  infinite expansions for the $H^{\rm eff}(\mathfrak{h})$ operator, Eq.~(\ref{duccexth}), are as follows:
\begin{itemize}
\item {\bf Rank of the commutator expansion.} All approximations introduced in past years are based on finite-rank approximations. In the most accurate approximations, first-, second-, and classes of third-rank commutators are included.
\item {\bf Source of the $\sigma_{\rm ext}(\mathfrak{h}$ amplitudes.}
As in the Eq.~(\ref{sigext}), the $\sigma_{\rm ext}(\mathfrak{h})$ are approximated in a  UCC way. In practical realizations considered so far, 
$T_{\rm ext}(\mathfrak{h})$ are extracted from the converged CCSD amplitudes.
\item {\bf Rank of many-body interactions in $H^{\rm eff}(\mathfrak{h})$.} Currently, downfolded Hamiltonians are constructed to include one- and two-body effective interactions. 
\item {\bf Perturbative consistency.} Many-body perturbation theory (MBPT) allows one to better balance the correlation effects in the expansion in Eq.~(\ref{duccexth}) (see extensive discussion of MBPT expansions for UCC theories in Refs.~\cite{unitary1,unitary2}). For perturbative "consistency" in some cases, we include Fock operator ($F_N$) dependent terms.
\end{itemize}
Various approximations for downfolded Hamiltonians discussed in Ref.~\cite{bauman2022coupled2c} are collected in Table~\ref{table_approx}.
\renewcommand{\tabcolsep}{0.2cm}
\begin{center}
\begin{table*}
  \centering
  \caption{Various approximations for the  $\bar{H}_{\rm ext}$ operator. Special notation is used to designate the perturbative structure of expansions employed. For example, $X^{(2)}$ and $X^{(3)}$ 
  terms on the right-hand side
designate parts of the operator $X$ correct up to the second and third order of MBPT. For the case of the natural orbitals, the same expressions are used with the full non-diagonal form of the Fock operator $F_N$. The normal product form of the electronic Hamiltonian $H_N$ is defined as $H_N=H-\langle\Phi|H|\Phi\rangle$.
}
  \begin{tabular}{lc} \hline \hline  \\
A(1) & $\bar{H}_{\rm ext}^{\rm A(1)} = H$ \\[0.3cm]
A(2) & $\bar{H}_{\rm ext}^{\rm A(2)} = H+[H_N,\sigma_{\rm ext}]^{(2)} + \frac{1}{2}[[F_N,\sigma_{\rm ext}],\sigma_{\rm ext}]^{(2)}$ \\[0.3cm]
A(3) & $\bar{H}_{\rm ext}^{\rm A(3)} = H+[H_N,\sigma_{\rm ext}] $ \\[0.3cm]
A(4) & $\bar{H}_{\rm ext}^{\rm A(4)} =H+[H_N,\sigma_{\rm ext}] +\frac{1}{2}[[F_N,\sigma_{\rm ext}],\sigma_{\rm ext}] $ \\[0.3cm]
A(5) & $\bar{H}_{\rm ext}^{\rm A(5)} = H+[H_N,\sigma_{\rm ext}]^{(3)} +\frac{1}{2}[[H_N,\sigma_{\rm ext}],\sigma_{\rm ext}]^{(3)}$ \\[0.2cm]
& $+\frac{1}{6} [[[F_N,\sigma_{\rm ext}],\sigma_{\rm ext}],\sigma_{\rm ext}]^{(3)}$ \\ [0.3cm]
A(6) & $\bar{H}_{\rm ext}^{\rm A(6)} = H+[H_N,\sigma_{\rm ext}]+\frac{1}{2}[[H_N,\sigma_{\rm ext}],\sigma_{\rm ext}]$ \\[0.3cm]
A(7) & $\bar{H}_{\rm ext}^{\rm A(7)} = H+[H_N,\sigma_{\rm ext}]+\frac{1}{2}[[H_N,\sigma_{\rm ext}],\sigma_{\rm ext}] $ \\[0.2cm]
& $+\frac{1}{6} [[[F_N,\sigma_{\rm ext}],\sigma_{\rm ext}],\sigma_{\rm ext}]$ \\[0.3cm]	
  \hline \hline 
  \end{tabular}
  \label{table_approx}
\end{table*}
\end{center}   
Obtaining Hermitian downfolded Hamiltonians, especially those including higher-rank commutators, is usually associated with the inclusion of hundreds or thousands of Hugenholtz-type diagrams, which requires developing and utilizing specialized symbolic tools to derive and efficiently implement the corresponding algebraic expressions. 

The Hermitian versions of the downfolded Hamiltonians have been integrated with various VQE and QPE solvers. The efficiency of these workflows has been illustrated in the example of bond-stretching processes for typical benchmark systems such as H$_2$, LiH, Li$_2$, N$_2$, H$_2$O, and C$_2$H$_4$ systems.\cite{metcalf2020resource,bauman2020variational,bauman2022coupled2c}
In all cases, downfolded Hamiltonians significantly improve the accuracy of diagonalization of the bare Hamiltonians in active spaces and provides results much closer to the exact (or nearly exact) results obtained when all orbitals are correlated.

\section{Conclusions}
The CC downfolding techniques are a relatively new tool to analyze/derive new properties of CC methods. One of the most appealing ones is the possibility of calculating CC energies 
through the diagonalization of the effective/downfolded Hamiltonians in broad classes of active spaces corresponding to standard CC approximations. These observations have been extended to the time-domain and quantum flows, which provide an alternative way to rigorously encapsulate the inherent sparsity/sparsities of quantum systems. 

Aside from interesting new properties of  CC methodology, the downfolding techniques have been used as a design principle to select classes of amplitudes in the vain of the Equivalence Theorem. These methods proved efficient in treating strong correlation effects in ground and excited states. 

The Hermitian form of the downfolding formalism is a promising extension for quantum computing applications. Due to the possibility of reducing the dimensionality of the quantum problem, it enables simulations for effective representations of Hamiltonians in basis sets that would be beyond the reach of current simulators and hardware if direct approaches are used. Since the current quantum algorithm, such as the VQE methodologies, can effectively handle only a relatively small number of wave function parameters, usually corresponding to the so-called static correlation effects, the active-space-driven downfolding is a potential tool to extend the area of applications to larger systems and larger basis sets. As a next frontier, our group is intensively developing quantum algorithms based on the unitary CC flow equations. These methods can traverse larger sub-spaces of Hilbert spaces than currently possible using modest quantum computing resources associated with the size of the maximum active space involved in the flow. This is a consequence of the fact that in the global representation of the quantum problem, requiring a full qubit register to assure the antisymmetry of the wave function, is replaced/approximated by flows (or computable reduced dimensionality eigenvalue problems) where global antisymmetry problems no longer exist. 

New computational paradigms associated with the emergence and broad utilization of machine learning techniques offer an exciting avenue for utilizing Hermitian CC downfolding to extract the analytical form of effective inter-electron interactions. These "phenomenological" interactions are ideal candidates to be integrated with the low-rank formulations such as Hartree--Fock, Density Functional Theory, and various types of multi-configurational self-consistent field methods as described in Ref.~\cite{bauman2022coupled}. 

\section{Acknowledgement} 
The main part of this  work was supported by  the Quantum Science Center (QSC), a National Quantum Information Science Research Center of the U.S. Department of Energy (DOE).
NPB and BP acknowledge the  support from  “Embedding Quantum Computing into Many-body Frameworks for Strongly Correlated Molecular and Materials Systems” project, which is funded by the U.S. Department of Energy (DOE), Office of Science, Office of Basic Energy Sciences, the Division of Chemical Sciences, Geosciences, and Biosciences.
All work was performed at Pacific Northwest National Laboratory (PNNL)
operated for the U.S.  Department of Energy by the Battelle Memorial Institute under Contract DE-AC06-76RLO-1830.
One of the authors of this review (KK) would like to express his deep gratitude to all his colleagues, friends, and mentors he had the honor to work with and learn from during his 
early "CC days."  Countless discussions with the quantum chemistry pioneers in Poland have inspired part of the presented material.


\end{document}